# Time Displaced Entanglement and Non-Linear Quantum Evolution

T. C. Ralph,
Department of Physics, University of Queensland, Brisbane 4072, QLD, Australia
(Dated: July 4, 2018)

We discuss time displaced entanglement, produced by taking one member of a Bell pair on a round trip at relativistic speeds, thus inducing a time-shift between the pair. We show that decoherence with respect to Bell measurements on the pair is predicted. We then study a teleportation protocol, using time displaced entanglement as its resource, in which a time-like loop is apparently formed. The result is non-unitary, non-linear evolution of the teleported state.

The apparent contradiction between the non-local correlations of entangled quantum systems and the local nature of relativity is normally resolved by the intrinsically probabilistic nature of quantum processes that prevent the extraordinary nature of the correlations being revealed without direct communication between the entangled parties. This is a satisfactory resolution when only inertial frames are considered [1]. However, the lack of a complete theoretical description of quantum processes in a relativistic framework leaves open the possibility of fundamentally new phenomena arising when non-inertial frames are involved.

Time-like curves, predicted by some solutions of classical relativity [2], allow a particle to follow a trajectory into its own past. Time-like curves can lead to acausal loops and hence paradoxes for classical systems. However, Deutsch has shown that the probabilistic nature of quantum mechanics avoids such paradoxes and leads to consistent solutions for quantum systems in the presence of time-like curves [3]. Although physically well-behaved, new phenomena is displayed, as the quantum systems exhibit non-unitary and non-linear evolution. This result is not only of academic interest as Bacon has shown that such evolutions could lead to a broader class of computable functions [4].

The fact that nature apparently allows such effects motivates a search for physical systems that might display them. We consider teleportation [5]. Teleportation allows the quantum state of a particle to be transferred to another particle without any direct interaction, using entanglement. We study a model of teleportation with a time-displaced entanglement resource. We begin by discussing a quite generic model, but consider a possible realization later. The model predicts that strong non-unitary and non-linear evolution can occur under certain conditions.

We consider two qubits initially at rest. We use the Schwinger model for angular momentum [6] to construct the following temporal mode representation of their logical states:

$$
\begin{aligned}
|0\rangle_t &= \int_{t-\Delta}^{t+\Delta} dt' F(t') \hat{a}_{t'}^\dagger |g\rangle_a |g\rangle_b \\
|1\rangle_t &= \int_{t-\Delta}^{t+\Delta} dt' F(t') \hat{b}_{t'}^\dagger |g\rangle_a |g\rangle_b
\end{aligned} \quad (1)
$$

where $\hat{a}_{t'}^\dagger$ and $\hat{b}_{t'}^\dagger$ are single time creation operators corresponding to two degrees of freedom of a simple harmonic oscillator, with corresponding ground-states $|g\rangle_a$ and $|g\rangle_b$. $F(t')$ is the temporal wave function of the qubit, defining its measurement and interaction bandwidths, and related by Fourier transform to the energy spectra of the qubit states. Notice that this representation requires that our qubits are broadband in energy. We have assumed for convenience that $F(t') = 0$ for $t' \geq |\Delta|$ thus allowing us to bound the limits of the integrals. Also for convenience we introduce an $nth$ clock-cycle time, $t_n$ such that $t_n - t_{n-1} > 2\Delta$. A measurement (or gate) at the $nth$ clock-cycle must be carried out between times $t_n - \Delta$ and $t_n + \Delta$ to definitely detect (or transform) the qubit state. The qubit states $|0\rangle_{t_n}$ and $|1\rangle_{t_n}$ form a complete basis for all the possible qubit states at the $nth$ clock-cycle. Finally, time evolution of an arbitrary qubit can be represented by:

$$
\begin{aligned}
\hat{U}(d)|\sigma\rangle_t &= \int_{t-\Delta}^{t+\Delta} dt' F(t')(\alpha\, \hat{a}_{t'+d}^\dagger |g\rangle_a |g\rangle_b \\
&\quad + \beta\, \hat{b}_{t'+d}^\dagger |g\rangle_a |g\rangle_b) \\
&= |\sigma\rangle_{t+d}
\end{aligned} \quad (2)
$$

Using this formalism we can write the following representation of an entangled qubit pair at the $nth$ clock-cycle:

$$
|\phi^+\rangle_{t_n} = |0\rangle_{t_n,1}|0\rangle_{t_n,2} + |1\rangle_{t_n,1}|1\rangle_{t_n,2} \quad (3)
$$

where we have introduced the additional subscripts, 1 and 2, to label different spatial locations of the two qubits. The details of the spatial wave-functions (assumed identical for the two qubits) are suppressed.

Initially the two qubits are in the same inertial frame. Now suppose qubit 1 is taken on a round-trip at high velocity relative to this rest frame. After the trip the qubit is returned to the initial inertial frame. We expect from classical relativity theory that the time evolution of the qubit taken on the round-trip will have slowed relative to the stay at home qubit, thus effectively travelling into the future. This can be modeled as a local application to qubit 1 of the time translation operator [7, 8]. For energy eigenstate qubits this would only produce a phase shift. However, for our broadband qubits it results in a shift in the proper time of qubit 1. Suppose the difference in proper time between the two qubits after the round trip,

$\tau$, is equal to one clock-cycle time, ie: $\tau = t_n - t_{n-1}$. The entanglement will still be expressed as occurring between the $n$th clock-cycle of each qubit. However, due to the time dilation, the $(n-1)$th clock-cycle of qubit 1 now occurs at $t_n$ according to clocks in the initial inertial frame. We are thus led to write the following description of the entangled state after the round-trip:

$$\begin{aligned}|\bar{\phi}^+\rangle &= |0\rangle_{t_n,1}|0\rangle_{(t_n-\tau),2} + |1\rangle_{t_n,1}|1\rangle_{(t_n-\tau),2} \\ &= |0\rangle_{t_n,1}|0\rangle_{t_{n-1},2} + |1\rangle_{t_n,1}|1\rangle_{t_{n-1},2}\end{aligned} \quad (4)$$

We refer to Eq.4 as time displaced entanglement and we will now study some unusual properties of this state. Previous authors have explored relativistic effects that arise whilst qubits are in different inertial frames [9] or noninertial frames [10]. However here, as both qubits are now again in the same inertial frame, we will simply apply non-relativistic quantum techniques to their analysis.

Notice first that local measurements on the individual qubits will not reveal any difference between Eq.4 and Eq.3. The time ordering of measurements does not affect the observed "Bell correlations" for the state of Eq.3, thus the internal rearrangement of time ordering that characterizes Eq.4 does not lead to an observable difference. However this is not true if we consider joint measurements. Consider the 2-mode projective measurement onto the basis spanned by the four Bell states:

$$\begin{aligned}|\phi^\pm\rangle &= |0\rangle_{t,1}|0\rangle_{t,2} \pm |1\rangle_{t,1}|1\rangle_{t,2} \\ |\psi^\pm\rangle &= |0\rangle_{t,1}|1\rangle_{t,2} \pm |1\rangle_{t,1}|0\rangle_{t,2}\end{aligned} \quad (5)$$

where for simplicity we now just write $t$ for the $n$th clock cycle time. Such a "Bell measurement" performed on the state of Eq.3 will return the result "$|\phi^+\rangle$" with unit probability. To model the measurement on the time displaced state we must consider the tensor product of the state Eq.4 with itself after evolution $\hat{U}(\tau)$:

$$\begin{aligned}(|0\rangle_{t,1}|0\rangle_{t-\tau,2} &+ |1\rangle_{t,1}|1\rangle_{t-\tau,2}) \\ &\otimes (|0\rangle_{t+\tau,1}|0\rangle_{t,2} + |1\rangle_{t+\tau,1}|1\rangle_{t,2})\end{aligned} \quad (6)$$

Notice that this tensor product is justified because the two states occupy non-overlapping Hilbert spaces. That is, the arbitrary states $|\sigma\rangle_{t_i,n}$ and $|\epsilon\rangle_{t_j,m}$ are normalized over distinct Hilbert spaces for $i \neq j$ and/or $n \neq m$. We wish to describe a Bell measurement occurring at the particular clock time, $t$, but Eq.6 contains a component describing qubit 1 at clock time $t_{n+1} = t + \tau$, i.e. after the measurement has occurred. Clearly the measurement will have removed this component, but will also have projected qubit 2 into some unknown state. We thus write:

$$\begin{aligned}(|0\rangle_{t,1}|0\rangle_{t-\tau,2} &+ |1\rangle_{t,1}|1\rangle_{t-\tau,2})c.c. \\ &\otimes (\gamma_{00}|0\rangle_{t,2}\langle 0|_{t,2} + \gamma_{11}|1\rangle_{t,2}\langle 1|_{t,2} \\ &+ \gamma_{01}|0\rangle_{t,2}\langle 1|_{t,2} + \gamma_{10}|1\rangle_{t,2}\langle 0|_{t,2})\end{aligned} \quad (7)$$

where the $\gamma's$ are parameters to be determined. A measurement on this state at time $t$ then projects qubit 2 into a particular state at time $t - \tau$ which must be consistent with (i.e. equal to) our unknown state. In this way the parameters can be found and hence the probabilities of various measurement outcomes. In particular a general solution for Bell state measurements is given by $\gamma_{00} = \gamma_{11} = 1/2, \gamma_{01} = \gamma_{10} = 0$, i.e. a completely mixed state. By substitution of this solution into Eq.7 it can easily be confirmed that all four Bell states can be detected with equal probabilities and that the solution for the projected state of qubit 2 is the completely mixed state as required (note we will return to this point in our concluding remarks).

Now let us study teleportation using the time displaced entanglement of Eq.4. Consider first Fig.1(a). As usual we assume Alice holds particle 1 and Bob holds particle 2. Alice makes a Bell measurement on particle 1 and an unknown qubit $\alpha|0\rangle_{t,3} + \beta|1\rangle_{t,3}$ which, for simplicity, we take to be in a pure state. Suppose Alice obtains the result "$|\phi^+\rangle$". It is easy to show that Bob's qubit is projected into the state $\alpha|0\rangle_{(t-\tau),2} + \beta|1\rangle_{(t-\tau),2}$. Notice that Bob receives the state before Alice sent it. The state has been teleported into the past. This apparent paradox has of course been previously discussed in the context of Alice and Bob being space-like separated [5, 11]. The resolution comes from realizing that other measurement results of Alice require bit-flip, phase-flip or bit-flip and phase-flip corrections which Bob has no way of knowing about until he "catches up" with Alice's measurement result. In the absence of this information Bob effectively holds a maximally mixed state.

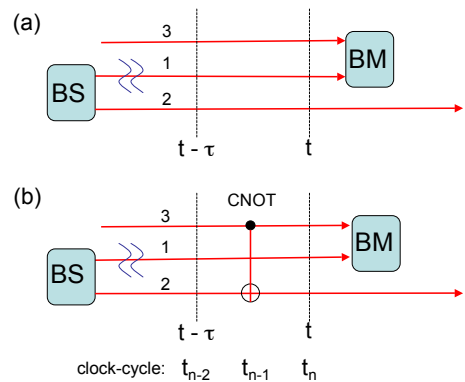

FIG. 1: Teleportation with time displaced entanglement. BS indicates Bell state production and BM indicates Bell state measurement. The wavy lines indicate the relativistic round trip. In (a), although mathematically the Bell measurement projects the state of the unknown qubit 3 (at time $t$) into the past of qubit 2 (at time $t-\tau$) there is no way of confirming this until the results of the Bell measurement become available, after time $t$. In (b) we perform an entangling gate between qubits 2 and 3 in the interval $t-\tau$ to $\tau$ that we expect may lead to more interesting effects.

Now consider Fig.1(b). For convenience we particularly now require that $\tau = t_n - t_{n-2}$, i.e. 2 clock cycles,

thereby allowing an interaction to occur between Bob's particle 2 and Alice's particle 3, at clock cycle $t_{n-1}$ (see Fig.1(b)). The situation is now more interesting. Even though we do not know at the time of the interaction the state of Bob's qubit, we are led to retrodict after Alice's Bell measurement (in particular if the result is $|\phi^+\rangle$) that the interaction occurred between the teleported qubit and its past self. Clearly this is an unusual situation and seems to involve a time-like curve. Importantly there is a free evolving particle that emerges from this interaction (qubit 2) which we might expect has suffered some interesting evoluton. In particular we assume a control-NOT (CNOT) gate is applied between the qubits with particle 3 as the control and particle 2 as the target. The CNOT transforms our qubit states in the following way: $|0\rangle_{t,3}|0\rangle_{t,2} \to |0\rangle_{t,3}|0\rangle_{t,2}$; $|0\rangle_{t,3}|1\rangle_{t,2} \to |0\rangle_{t,3}|1\rangle_{t,2}$; $|1\rangle_{t,3}|0\rangle_{t,2} \to |1\rangle_{t,3}|1\rangle_{t,2}$; $|1\rangle_{t,3}|1\rangle_{t,2} \to |1\rangle_{t,3}|0\rangle_{t,2}$. We now seek a solution for the state of qubit 2 after this interaction.

We proceed by considering the state of qubits 2 and 3 at time $t$, just before the Bell measurement is made. We write this state in the completely general form:

$$\rho_{3,2}(t) = \sum_{i,j,k,l=0,1} \gamma_{ijkl} |i\rangle_{t,3} |j\rangle_{t,2} \langle k|_{t,2} \langle l|_{t,3} \quad (8)$$

where again the $\gamma$'s are parameters to be determined. We now make the Bell measurement between qubit 1 and qubit 3, with the state of qubit 1 (and the past state of qubit 2) described by Eq.4. Suppose the result is $|\phi^+\rangle$, the state of the system is now:

$$\rho_2(t,t') = \sum_{i,j,k,l=0,1} \gamma_{ijkl} |i\rangle_{t',2} |j\rangle_{t,2} \langle k|_{t,2} \langle l|_{t',2} \quad (9)$$

describing the state of qubit 2 at clock times $t = t_n$ and $t' = t - \tau = t_{n-2}$. We tensor this state with the state of qubit 3 at time $t'$: $\alpha|0\rangle_{t',3} + \beta|1\rangle_{t',3}$. We now evolve this whole system forward in time by $\tau$, through the CNOT gate where the past states of qubit 3 and qubit 2 interact. We now have a state describing the states of qubits 3 and 2 at time $t$, and qubit 2 at time $t + \tau$. We are only interested in the state at time $t$ so we trace out the future component leaving:

$$\begin{aligned}\rho_{3,2}(t) &= (\alpha^2 |0\rangle_{t,3}|0\rangle_{t,2}\langle 0|_{t,2}\langle 0|_{t,3} + \beta^2 |1\rangle_{t,3}|1\rangle_{t,2}\langle 1|_{t,2}\langle 1|_{t,3} \\ &+ \alpha\beta(|0\rangle_{t,3}|0\rangle_{t,2}\langle 1|_{t,2}\langle 1|_{t,3} + |1\rangle_{t,3}|1\rangle_{t,2}\langle 1|_{t,2}\langle 1|_{t,3})) \\ &\quad \times (\gamma_{0000} + \gamma_{0110}) \\ &+ (\alpha^2 |0\rangle_{t,3}|1\rangle_{t,2}\langle 1|_{t,2}\langle 0|_{t,3} + \beta^2 |0\rangle_{t,3}|1\rangle_{t,2}\langle 1|_{t,2}\langle 0|_{t,3} \\ &+ \alpha\beta(|0\rangle_{t,3}|1\rangle_{t,2}\langle 0|_{t,2}\langle 1|_{t,3} + |1\rangle_{t,3}|0\rangle_{t,2}\langle 1|_{t,2}\langle 0|_{t,3})) \\ &\quad \times (\gamma_{1001} + \gamma_{1111}) \\ &+ (\alpha^2 |0\rangle_{t,3}|0\rangle_{t,2}\langle 1|_{t,2}\langle 0|_{t,3} + \beta^2 |1\rangle_{t,3}|1\rangle_{t,2}\langle 0|_{t,2}\langle 1|_{t,3} \\ &+ \alpha\beta(|0\rangle_{t,3}|0\rangle_{t,2}\langle 0|_{t,2}\langle 1|_{t,3} + |1\rangle_{t,3}|1\rangle_{t,2}\langle 1|_{t,2}\langle 0|_{t,3})) \\ &\quad \times (\gamma_{0001} + \gamma_{0111}) \\ &+ (\alpha^2 |0\rangle_{t,3}|1\rangle_{t,2}\langle 0|_{t,2}\langle 0|_{t,3} + \beta^2 |1\rangle_{t,3}|0\rangle_{t,2}\langle 1|_{t,2}\langle 1|_{t,3} \\ &+ \alpha\beta(|0\rangle_{t,3}|1\rangle_{t,2}\langle 1|_{t,2}\langle 1|_{t,3} + |1\rangle_{t,3}|0\rangle_{t,2}\langle 0|_{t,2}\langle 0|_{t,3})) \\ &\quad \times (\gamma_{1000} + \gamma_{1110})\end{aligned} \quad (10)$$

If we are to have consistent evolution then Eq.8 and Eq.10 should be the same. By equating terms and insisting that the final state be normalized we arrive at the unique solutions:

$$\begin{aligned}\gamma_{0000} &= \alpha^4, \quad \gamma_{0110} = \gamma_{1111} = \alpha^2\beta^2 \\ \gamma_{1001} &= \beta^4, \quad \gamma_{0011} = \gamma_{1100} = \alpha^3\beta \\ \gamma_{0101} &= \gamma_{1010} = \alpha\beta^3\end{aligned} \quad (11)$$

with all other $\gamma$'s equal to zero. Substituting back into Eq.10 and tracing over qubit 3 we obtain the state of the free-propagating qubit 2:

$$\rho_2 = (\alpha^4 + \beta^4)|0\rangle\langle 0| + 2\alpha^2\beta^2|1\rangle\langle 1| \quad (12)$$

Eq.12 is the second and major result of this paper. It features non-linear non-unitary evolution of the input qubit to the output qubit.

So far we have only considered the solution when the result "$|\phi^+\rangle$", is eventually obtained by the Bell measurement. Solutions for the other measurement results can be obtained in a similar way. It is found that the solution for the output is unchanged for the result $|\phi^-\rangle$, whilst for $|\psi^\pm\rangle$, the result is a bit flipped version of Eq.12, which can be corrected when the result is known. Notice that, as required for preservation of causality, without the knowledge of the Bell measurement result, the qubit is in a maximally mixed state.

Eq.12 is the same evolution as that predicted by Bacon [4] in the context of a CNOT + SWAP interaction between a free evolving qubit and a closed time-like curve generated by a quantum worm-hole. In our case, in effect, the combination of the time-displaced entanglement and the Bell measurement form a closed time-like curve. We thus refer to our process as *time-loop teleportation*. The connection between closed time-like curves and the Bell state generation and measurement sequence has been discussed from an interpretational point of view by Pegg [12].

The highly unusual properties of our system can be illustrated by considering how the trace distance $D = Tr|\rho_A - \rho_B|$ between the logical zero state and a real superposition state (i.e. $\alpha, \beta$ real) changes due to the evolution. At the input $D = 2\beta^2$. Suppose $D_a$ is the trace distance after evolution. Normally it is always true that $D_a \leq D$, meaning that the distinguishability between two quantum states cannot be increased. However from Eq.12 we have $D_a = 4(\beta^2 - \beta^4)$. As shown in Fig.2, distinguishability is increased in the region $0 < \beta^2 < 1/2$. Notice, though, that outside this region distinguishability is reduced. Indeed, on average, intergrating around the real great circle, distinguishability is reduced. Nevertheless, Bacon [4] has shown that significantly increased computing power is implied by evolution of this kind.

Up to this point our description of the qubit has been rather abstract. We now introduce a particular physical implementation as an example of our formalism. Consider a single photon that has been trapped in a loop of fibre (or some other suitable delay line).We assume the fibre is loss-less and dispersion-less. Qubit values can be defined by the polarization of the photon, taking, for example, horizontal to be logical zero and vertical to be logical one. We let the round-trip time of the photon in the fibre be $\tau$, which defines the clock-cycle. Each clock-cycle the qubit can be described by Eq.1 where now $\hat{a}^\dagger_{t'}$ and $\hat{b}^\dagger_{t'}$ represent single time creation operators for the two optical polarization modes, with the ground-states $|g\rangle_a$ and $|g\rangle_b$ now corresponding to the respective electromagnetic vacuum modes. $F(t')$ is the spatio-temporal wave packet of the photon. As before we require $2\Delta < \tau$ to ensure that the photon wave-packet is shorter than the clock-cycle. We assume that it is possible to make single qubit rotations on the photons and that at particular clock-cycles it is possible to have photons in different fibre loops interact through a non-linear medium capable of inducing a CNOT gate [13].Finally we assume it is possible to release a photon from its fibre at some particular clock-cycle and measure its polarization state. Using the CNOT gate, entanglement can be produced between the fibre photon qubits, one of which can then be taken on the round-trip. The CNOT gate can also be used to make Bell-measurements after the round-trip.

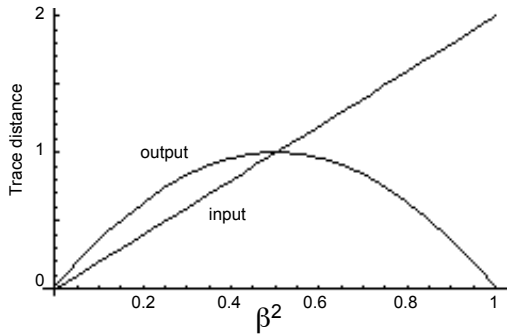

FIG. 2: Plot of the trace distance between the logical zero state and an arbitrary real superposition, at the input and after the time-loop teleportation. For the region $\beta^2 < 0.5$ we have the extraordinary result that the trace distance is increased. Notice, though that for $\beta^2 > 0.5$ the trace distance is decreased. On average distinguishability is decreased.

We have described an interaction which combines quantum entanglement with relativistic effects. We have shown that the resulting evolution lies outside the realm of standard quantum mechanics. In particular non-linear state evolution is predicted. However, the general class of evolutions possible from this effect do not lead to paradoxes or violations of causality [3]. Indeed our physical formulation removes two ad hoc assumptions made in the more abstract previous work. Firstly, it was necessary to restrict the types of interactions allowed with the time-like loop. In particular measurement or preparation of the closed time-like qubits was not allowed [4]. However here, no restrictions are needed as no information can be extracted from the teleported state until the results of the Bell measurements are known. Secondly, under certain circumstances multiple self-consistent solutions could arise in the previous work. Consider a closed loop with no outside interaction. Any solution for the trapped qubit is allowed. Certain interactions with a free evolving qubit (for example a control-sign gate [4]) do not change this situation, such that the solutions for the free evolving qubit also seem unconstrained and some further assumptions must be made in order to pick a particular solution. The equivalent situation in our model is the Bell state measurement of the time displaced entanglement. Consider Eq.7. Suppose the result "$|\phi^+\rangle$" is obtained. It can be seen that any solution for the $\gamma$'s is allowed. However this is a singular solution. If we consider stability of the result against small perturbations away from the ideal measurement basis we find that the totally mixed solution is unique for all Bell state results, thus removing the ambiguity.

The key assumption leading to Eq.4 is that after the round-trip, with both qubits again in the same inertial frame, the only effect on the travelling qubit is the classically predicted change in proper time. Although this proposal would obviously be demanding to test experimentally it does not seem beyond the realm of horizon technology, with maintenance of coherence during the round-trip the likely biggest hurdle.

This investigation was motivated by a seminar on time in quantum mechanics given by David Pegg. I thank P.P.Rohde, C.M.Savage, K.Pregnell and G.J.Milburn for helpful discussions. This work was supported by the Australian Research Council.

[1] A.Stefanov, H.Zbinden, N.Gisin, and A.Suarez Phys. Rev. Lett. **88**, 120404 (2002).
[2] M.S.Morris, K.P.Thorne and U.Yurtsever, Phys.Rev.Lett. **61**, 1446 (1988).
[3] D.Deutsch, Phys.Rev.D **44**, 3197 (1991).
[4] D.Bacon, Phys.Rev.A, **70** 032309 (2004).
[5] C.H.Bennett, G.Brassard, C.Crepeau, R.Jozsa, A.Peres, W.K.Wootters, Phys.Rev.Lett. **70**, 1895 (1993).
[6] J.J.Sakurai, *Modern Quantum Mechanics*, (Addison-Wesley, Reading 1994).
[7] Y.Aharonov, J.Anandan, S.Popescu and L.Vaidman, Phys.Rev.Lett. **64** 2965 (1990); L.Vaidman, Found.Phys. **21** 947 (1991).
[8] S.L.Braunstein, C.M. Caves and G.J.Milburn, Annals of Physics, **247**, 135(1996).
[9] A.Peres, P.F.Scudo and D.R.Terno, Phys.Rev.Lett.

**88** 230402 (2002); P.L.Alsing and G.J.Milburn, Quant.Imf.Comp. **2**, 487 (2002); J.Pachos and E.Solano, Quant.Imf.Comp. **2**, 115 (2003); R.M.Gingrich and C.Adami, Phys.Rev.Lett. **89** 270402 (2002).
[10] W.G.Unruh, Phys.Rev.D **14**, 870 (1976); P.Kok and U.Yurtsever, Phys.Rev.D **68** 085006 (2003); P.M.Alsing and G.J.Milburn, Phys.Rev.Lett, **91** 180404 (2003); I.Fuentes-Schuller and R.B.Mann, Phys.Rev.Lett. **95**, 120404 (2005).
[11] C.Brukner, J-W.Pan, C.Simon, G.Weihs, and A.Zeilinger Phys. Rev. A **67**, 034304 (2003).
[12] D.T.Pegg, in *Time's Arrows, Quantum Measurement and Superluminal Behaviour*, edited by D.Mugnai et al (Consiglio Nazionale delle Richerche, Roma, 2001) p.113, quant-ph/0506141.
[13] G.J.Milburn, Phys.Rev.Lett. **62**, 2124 (1989).